\documentclass[aps,prl,floatfix,twocolumn]{revtex4}
\usepackage{graphicx}

\begin{document}

\title{Do all states undergo sudden death of entanglement at finite temperature? \vspace{-2mm}} 

\author{A. Al-Qasimi, D. F. V. James}
\affiliation{Department of Physics, University of Toronto, Toronto ON M5S 1A7, Canada.
\vspace{-2mm}}

\begin{abstract} \noindent 
In this paper we consider the decay of quantum entanglement, quantified by the concurrence, of a pair of two-level systems each of which is interacting with a reservoir at finite temperature T.  For a broad class of initially entangled states, we demonstrate that the system always becomes disentangled in a finite time i.e. ``entanglement sudden death'' (ESD) occurs. This class includes \emph{all} states which previously had been found to have long-lived entanglement in zero temperature reservoirs. Our general result is illustrated by an example.
\end{abstract}

\maketitle

\noindent 
In the past few years there has been considerable interest in the properties of entangled quantum systems.  Spurred on by the emergence of compelling applications in quantum information processing, useful methods by which the entanglement of quantum systems can be established and characterized have emerged.  Perhaps the most impactful to date has been the simple procedure derived by Wootters \cite{Wootters} for quantifying entanglement for an arbitrary mixed state of a pair of two-level systems.  This has provided a very useful tool for measurement of experimental quantum states \cite{DFVJ1} and is today commonly used in assessing the capabilities of emerging quantum technologies.
Building on Wootter's work, recently Yu and Eberly \cite{YuEberly1} investigated the time evolution of entanglement (quantified using the concurrence) of a bipartite qubit system undergoing various modes of decoherence.  Remarkably, they found that, even when there is no interaction, (either directly or through a correlated environment), there are certain states whose entanglement decays exponentially with time, while for other closely related states, the entanglement vanishes completly in a finite time.  This ``entanglement sudden death" (ESD) is an intriguing and potentially very important discovery.  Since the first theoretical demonstration of ESD, further investigations of different systems have been made by various groups \cite{YE,YYE,ASH,MAA,CLY,YE2,YYE2,Lastra,Zhang,VV}. Extending Yu and Eberly's model by considering correlated reservoirs and interactions \cite{YYE,YYE2,VV,Zhang,CLY}, it was shown that entanglement may be created by spontaneous emission (something which has been known for some time \cite{Almut} in a different context). The ESD has also been predicted for more complicated systems using other entanglement measures \cite{MMC,Isabel}, and an attempt to give a geometric interpretation for the phenomena of ESD has also been made \cite{Cunha}. Very recently, experimental studies have also been carried out to demostrate ESD, using carefully engineered interactions between systems and environments: Sudden death has been observed both in photons \cite{SDexp} and in atomic ensembles \cite{Kimble}.

From the technological point of view, states which exhibit exponential decay of entanglement, and therefore retain some vestige of this all-important correlation for long periods, are of great interest.  Thus it is important to identify precisely in what circumstances ESD will occur.  In this paper, we consider qubits in finite temperature reservoirs: instead of the energy of the qubits being lost via spontaneous decay to the environment, now additionally the reservoirs can cause excitation of the qubits.  For a broad class of mixed quantum states, which includes all of the states studied by Yu and Eberly and others in connection with this problem, we demonstrate that {\em all states undergo sudden death of entanglement at finite temperature.}  

As in \cite{YuEberly1}, we study a system of two qubits initially entangled and interacting with uncorrelated reservoirs. However, unlike \cite{YuEberly1}, in which the system is studied at T=0, we include the effects of heat in our system (Fig.~\ref{fig:figure1}). Here the dynamics of the density matrix $\hat{\rho}$ describing the two qubits is given by:

\begin{figure}[!b]
\vspace{-3mm}
\includegraphics[width=0.8 \columnwidth]{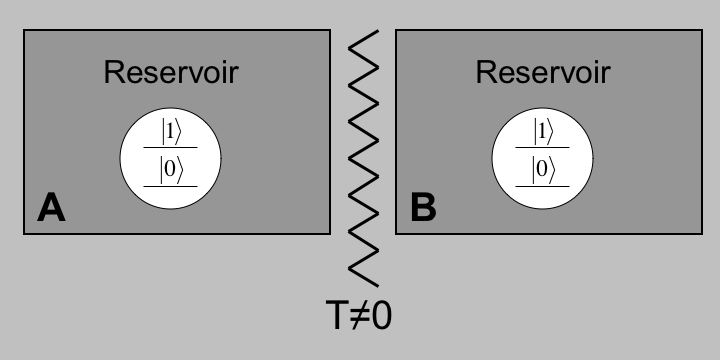}
\vspace{-2mm}
\caption{\textbf{Disentanglement by spontaneous emission of a two-qubit system in a heat bath.} The reservoir is modelled by different harmonic oscillator modes. Each qubit, here depicted by a two-level atom, interacts with its reservoir. The only interaction between the qubits that ever exists is the one that leads to their entanglement at time t =0. Following this, however, the only interaction that remains is that with the corresponding reservoir. This leads to decoherence, which causes the qubits to disentangle. Here we include the effect of heat by studying the system at T $\neq$ 0.}
\label{fig:figure1}
\end{figure}

\begin{equation}
\frac{\partial\hat{\rho}}{\partial t}=\frac{1}{i\hbar}[\hat{H},\hat{\rho}]+L_{1}[\hat{\rho}]+L_{2}[\hat{\rho}],
\label{B}
\end{equation}

\noindent where $[H,\rho]$ is the unitary part of the evolution (which we shall ignore as it has no effect on our study decoherence). The Liouvillian of the ith qubit is given by:
\begin{eqnarray}
& & L_{i}[\hat{\rho}]=\frac{(\bar{n}+1)\Gamma}{2}\left\{\left[\hat{\sigma}^{i}_{-}, \hat{\rho}\hat{\sigma}^{i}_{+}\right]+\left[\hat{\sigma}^{i}_{-}\hat{\rho}, \hat{\sigma}^{i}_{+}\right]\right\} \nonumber \\
& & \ \ \ \ \ \mbox{} + \frac{\bar{n}\Gamma}{2}\left\{\left[\hat{\sigma}^{i}_{+},
\hat{\rho}\hat{\sigma}^{i}_{-}\right]+\left[\hat{\sigma}^{i}_{+}\hat{\rho}, \hat{\sigma}^{i}_{-}\right]\right\},
\label{C} 
\end{eqnarray}

\noindent where $\Gamma$ is the spontaneous decay rate of the qubits, $\hat{\sigma}^{i}_{+}=(\left|1\right\rangle\left\langle 0\right|)_{i}$, and $\hat{\sigma}^{i}_{-}=(\left|0\right\rangle\left\langle 1\right|)_{i}$, where the index i (=1,2) denotes the qubits. The first term on the right hand side of (\ref{C}) corresponds to the depopulation of the atoms, while the second term describes the decay caused by the finite temperature; and $\bar{n}$ is the mean occupation number of the reservoir (assumed to be the same for both).

We assume that our system is initially an``X-state`` described by the following density matrix:
\begin{equation}
\hat{\rho}(t)=
\left(
\begin{array}{cccc} a(t) & 0 & 0 & w(t)  
\\             	   	0 & b(t) & z(t) & 0  
\\									0 & z^{\ast}(t) & c(t) & 0  
\\									w^{\ast}(t) & 0 & 0 & d(t)  
\end{array}
\right ).
\label{E}
\end{equation}

\noindent Such states are general enough to include states such as the Werner states, the maximally entangled mixed states (MEMS) \cite{MEMS}, the Bell States, and what we will refer to as the $\hat{\rho}_{YE}$ states, studied in \cite{YuEberly1} and which will be described later.

Substituting (\ref{E}) into (\ref{B}), the Master equation of our system, we obtain the following first order coupled differential equations:

\begin{eqnarray}
& & \dot{a}(t)=\Gamma \left[-2(\bar{n}+1)a(t)+b(t)\bar{n}+c(t)\bar{n}\right], \nonumber \\
& & \dot{b}(t)=\Gamma \left[(\bar{n}+1)a(t)-(2\bar{n}+1)b(t)+\bar{n}d(t)\right], \nonumber\\
& & \dot{c}(t)=\Gamma \left[(\bar{n}+1)a(t)-(2\bar{n}+1)c(t)+\bar{n}d(t)\right],\nonumber\\
& & \dot{d}(t)=\Gamma \left[(\bar{n}+1)b(t)+(\bar{n}+1)c(t)-2\bar{n}d(t)\right], \nonumber\\
& & \dot{z}(t)=\Gamma \left[-(2\bar{n}+1)z(t)\right], \nonumber\\
& & \dot{w}(t)=\Gamma \left[-(2\bar{n}+1)w(t)\right]. \nonumber\\
\label{F}
\end{eqnarray}

\noindent These may be solved to yield the following expressions:

\begin{eqnarray}
a(t)&&= \frac{1}{(2\bar{n}+1)^2}\left\{\bar{n}^{2}\right. \nonumber \\
& & \left.+[2(a_{0}-d_{0})\bar{n}^{2}+(a_{0}-d_{0}+1)\bar{n}]X \right. \nonumber \\
& & \left. +[(2a_{0}+2d_{0}-1)\bar{n}^{2}+(3a_{0}+d_{0}-1)\bar{n}+a_{0}]X^{2}\right\}, \nonumber \\
b(t)&&= \frac{1}{(2\bar{n}+1)^2}\left\{\bar{n}(\bar{n}+1) \right. \nonumber \\
& & \left. -[2(a_{0}+2c_{0}+d_{0}-1)\bar{n}^{2}+(a_{0}+4c_{0}+3d_{0}-2)\bar{n}\right. \nonumber \\
& & \left. +(c_{0}+d_{0}-1)]X \right. \nonumber \\
& & \left.-[(2a_{0}+2d_{0}-1)\bar{n}^{2} +(3a_{0}+d_{0}-1)\bar{n}+a_{0}]X^{2}\right\},\nonumber \\
c(t)&&= \frac{1}{(2\bar{n}+1)^2}\left\{ \bar{n}(\bar{n}+1) \right. \nonumber \\
& & \left. +[2(a_{0}+2c_{0}+d_{0}-1)\bar{n}^{2}+(3a_{0}+4c_{0}+d_{0}-2)\bar{n}]X \right. \nonumber \\
& & \left. -[(2a_{0}+2d_{0}-1)\bar{n}^{2}+(3a_{0}+d_{0}-1)\bar{n}+a_{0}]X^{2}\right\},\nonumber \\
d(t)&&= \frac{1}{(2\bar{n}+1)^2}\left\{(\bar{n}+1)^{2}\right. \nonumber \\
& & \left.-(\bar{n}+1)[2\bar{n}(a_{0}-d_{0})+(a_{0}-d_{0}+1)]X \right. \nonumber \\
& & \left.+[(2a_{0}+2d_{0}-1)\bar{n}^{2}+(3a_{0}+d_{0}-1)\bar{n}+a_{0}]X^{2} \right\}, \nonumber \\
w(t)&&= w_{0}X,\nonumber \\
z(t)&&= z_{0}X,\nonumber \\
\label{L}
\end{eqnarray}

\noindent where $X=e^{-\Gamma(2\bar{n}+1)t}$, $a_{0}$ =a(0), etc.

Using Wootters' formula \cite{Wootters}, the concurrence for a state of the form given by (\ref{E}) is:

\begin{equation}
C=2\, \mbox{max}\left\{0,|z(t)|-\sqrt{a(t)d(t)},|w(t)|-\sqrt{b(t)c(t)}\right\}.
\label{N}
\end{equation}

\noindent This implies that the disentanglement time will be the largest positive solutions of the following equations:

\begin{equation}
|z(t)|-\sqrt{a(t)d(t)}=0, |w(t)|-\sqrt{b(t)c(t)}=0.
\label{O}
\end{equation}

\noindent Multiplying both equations in (\ref{O}) by the positive quantities $|z(t)|+\sqrt{a(t)d(t)}$ and $|w(t)|+\sqrt{b(t)c(t)}$, respectively, yields:

\begin{equation}
|z(t)|^{2}-a(t)d(t)=0,\ |w(t)|^{2}-b(t)c(t)=0.
\label{O'}
\end{equation}

\noindent Substituting from equation (\ref{O'}) gives two quartic equations in X, which we will use in the proof of our main result.

The quantity X is the time-dependent parameter that we use to monitor the evolution of entanglement in the system. Notice that at t=0, X=1, and that at t=$\infty$, X=0. Physically meaningful values for X are, therefore, between 0 and 1. Asymptotic decay of entanglement implies a solution at X=0. However, if the entanglement of the system decays in a finite time (ESD), the solution of (\ref{O'}) must lie in the range $0 < X <1$.

Both equations in (\ref{O'}) are polynomial equations in X and continuous. At X=0, these equations take the following value:
\begin{equation}
\frac{-\bar{n}^{2}(\bar{n}^{2}+2\bar{n}+1)}{(2\bar{n}+1)^{4}}.
\label{T}
\end{equation}

\begin{figure}[!b]
\vspace{-3mm}
\includegraphics[width=0.8 \columnwidth]{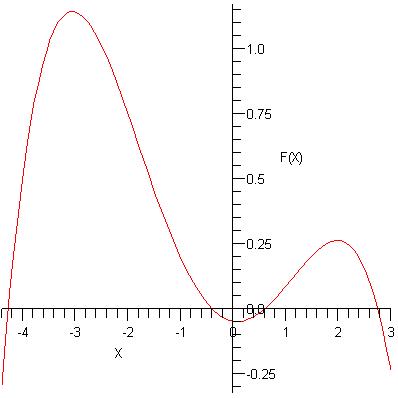}
\vspace{-2mm}
\caption{\textbf{Plot of F(X) vs X}. This is the plot of the first quartic equation in (\ref{O'}) for n =0.8, a =0.1, d = 0.05, z =0.3.}
\label{fig:figure2}
\end{figure}

\noindent Notice that since $\bar{n}$ is a positive quantity, (\ref{T}) is negative. On the other hand, if we evaluate (\ref{O'}) for X=1, which corresponds to t=0, we obtain $|z|^{2}-ad$ and $|w|^{2}-bc$ for the first and second equations, respectively. At least one of these quantities has to be positive if we assume that our systems are initially entangled, which is the case here. The fact that the quartic equations are continuous and have a negative value at X=0 and a positive one at X=1 implies that they have at least one root in our interval of interest $0 < X <1$. See Fig.~\ref{fig:figure2}.

For finite $\bar{n}$, and therefore for finite T, the constant term (\ref{T}) is always finite and nonzero for $\bar{n}>0$. Hence, there is no X=0 solution; i.e., no asymptotic decay. Equation (\ref{O'}) has at least one solution in the range $0 < X <1$. implying that the entanglement falls to zero in a \emph{finite time.}

As an example, let us consider a special case when w(t)=0. In this case, the only quartic equation that has to be satisfied for C=0 is the first one in (\ref{O'}). This equation can be easily solved with appropriate reparametrization. The four solutions for X are given by:
\begin{eqnarray}
& & X=(r+s)\pm\sqrt{(r+s)^{2}+t^{2}}, \nonumber \\
& & X=(r- s)\pm\sqrt{(r- s)^{2}+t^{2}}.
\label{P}
\end{eqnarray}

\noindent where
\begin{eqnarray*}
& & r=\frac{1+(a_{0}-d_{0})(2\bar{n}+1)}{4\left\{\left(2\bar{n}+1\right)\left[a_{0}(\bar{n}+1)+d_{0}\bar{n}\right]-\bar{n}(\bar{n}+1)\right\}},\\
& & s=\frac{(2\bar{n}+1)^{2}\sqrt{(1-a_{0}-d_{0})^{2}+4(z^{2}-a_{0}d_{0})}}{4\left\{\left(2\bar{n}+1\right)\left[a_{0}(\bar{n}+1)+d_{0}\bar{n}\right]-\bar{n}(\bar{n}+1)\right\}},\\
& & t=\sqrt{\frac{\bar{n}(\bar{n}+1)}{\left\{\left(2\bar{n}+1\right)\left[a_{0}(\bar{n}+1)+d_{0}\bar{n}\right]-\bar{n}(\bar{n}+1)\right\}}}.
\end{eqnarray*}

Further, following (\ref{E}), let us consider states of the form:

\begin{equation}
\hat{\rho}_{YE}=
\frac{1}{3}\left(
\begin{array}{cccc} \alpha & 0 & 0 & 0  
\\             	   	0 & 1 & 1 & 0  
\\									0 & 1 & 1 & 0  
\\									0 & 0 & 0 & 1-\alpha  
\end{array}
\right ).
\label{U}
\end{equation}

\begin{figure}[!b]
\vspace{-3mm}
\includegraphics[width=0.8 \columnwidth]{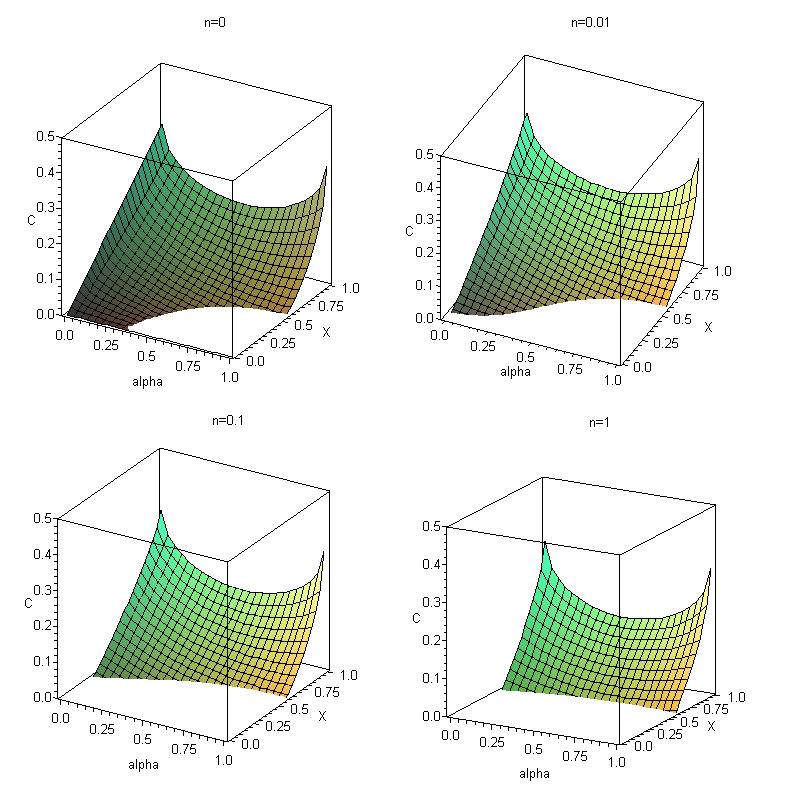}
\vspace{-2mm}
\caption{\textbf{Plot of concurrence (C) vs X vs a.} C=0 corresponds to no entanglement. X=1 corresponds to t=0, while X=1 corresponds to $t=\infty$. Notice that as soon as n becomes finite, for all values of $\alpha$, C becomes zero at $X < 0$; i.e., entanglement decays in a finite time. As n becomes bigger, all states disentangle at approximately X=0.5.}
\label{fig:figure3}
\end{figure}

\noindent In \cite{YuEberly1}, Yu and Eberly have shown that for $0 \leq \alpha \leq\frac{1}{3} $, the entanglement of this state is long-lived at zero temperature. Here, we will illustrate that as soon as $\bar{n}$ becomes finite, the range vanishes and there is no long-lived entanglement for any value of $\alpha$. The plots in Fig.~\ref{fig:figure3} illustrate this. As $\bar{n}$ increases, the range of values for $\alpha$ in which entanglement is long-lived diminishes until all values disappear.

In this paper, we presented a proof that in two-qubit systems interacting with uncorrelated reservoirs and described by X-states, ESD always occurs at any finite temperature. Although X-states are quite general states, they are not the most general ones. Thus the next question to ask is: do \emph{all} states exhibit ESD? In other words, for a state described by a density matrix, without any zero elements, is entanglement still lost in a finite time? This question is not straight forward. One has to find the equation for concurrence in that case, determine its order, study the properties of its coefficient, and from there maybe be able to comment on the nature of the roots, and, therefore, be able to predict what will happen in the actual physical system. However, our results demonstrate that in any finite temperature reservoir, all states that have been shown to be long-lived in a zero-temperature bath, will undergo sudden death.

We would like to thank Joseph Eberly, Ren$\rm{\acute{e}}$ Stock, and Bill Munro for valuable discussions. This work was supported by NSERC.

\end{document}